\documentclass{llncs}

\usepackage{amsfonts}
\usepackage{amsmath}
\usepackage{amssymb}

\usepackage{color}

\usepackage{tikz}
\usepackage{pgfpages}

\usetikzlibrary{patterns}

\newtheorem{conclusion}[theorem]{Conclusion}

\newcommand{\sign}{\text{sign}}

\tikzstyle{terminal}=[draw,fill=black, inner sep=1.5pt]
\tikzstyle{Steinerpoint}=[circle,draw,fill=white, inner sep=1pt]

\begin{document}

\title{The rectilinear Steiner tree problem with given topology and length
restrictions}

\author{Jens Ma{\ss}berg}
\institute{Institut f\"ur Optimierung und Operations
Research,
Universit\"at Ulm, jens.massberg@uni-ulm.de}

\maketitle

\textbf{Keywords:} Steiner trees with given topology, rectilinear Steiner trees,
dynamic programming, totally unimodular, shallow light Steiner trees

\begin{abstract}
 We consider the problem of embedding the Steiner points of a Steiner tree with
given topology into the rectilinear plane. 
Thereby, the length of the path between a distinguished terminal and each other
terminal must not exceed given length restrictions. 
We want to minimize the total length of the tree.
 
The problem can be formulated as a linear program and therefore it is solvable
in polynomial time.
In this paper we analyze the structure of feasible embeddings and give a
combinatorial polynomial time algorithm for the problem.
Our algorithm combines a dynamic programming approach and binary
search and relies on the total unimodularity of a matrix appearing in a sub-problem.
\end{abstract}

\section{Introduction}



The {\sc Rectilinear Steiner Tree Problem With Given
Topology And Length Restrictions} can be stated as follows. The
input $(S,T,r,p,l)$
consists of a set of terminals $T$ with positions $p:T\rightarrow \mathbb{R}^2$,
a tree $S$ with $T\subseteq V(S)$, a distinguished terminal $r\in T$ - called
the
\emph{root} of the tree -
and length restrictions $l_{t}\in \mathbb{R}_{\geq 0}$ for all $t\in T$.

The task is to find an embedding $\pi:V(S)\to \mathbb{R}^2$ of the vertices of
the tree into the plane
with $\pi(t)=p(t)$ for all $t\in T$, such that
for all $t\in T$ the length $d_\pi(t)$ of the unique path  from $r$
to $t$ in $S$ with edge set $E_S[r,t]$ has length at most $l_t$, that is,
\begin{equation}
 d_\pi(t)=\sum_{\{v,w\}\in E_S[r,t]} ||\pi(v)-\pi(w)||_1 \leq l_t
\end{equation}
and the total length 
\begin{equation}
  c(\pi):= \sum_{\{v,w\}\in E(S)} ||\pi(v)-\pi(w) ||_1 
\end{equation}
of the tree is minimized.
The tree $S$ is called \emph{Steiner tree} and the vertices in $V(S)\setminus S$
\emph{Steiner points}.
Throughout this paper we assume w.l.o.g. that the root is placed at the origin,
that is, $p(r)=(0,0)$. By adding Steiner points and edges of length zero we can
assume that the terminals are leaves of $S$ and that all Steiner points have
degree $3$.
Moreover, we denote by $\pi_x(v)$ and $\pi_y(v)$ the $x$- and
$y$-coordinate, respectively, of $\pi(v)$ for an
embedding $\pi$ and a vertex $v\in V(S)$.

A further generalization of the problem is to extend it to other metrics or to
consider length restrictions between any pair of terminals. In this paper we
restrict ourselves to the $\ell_1$ metric and length restrictions between one
distinguished vertex and all other terminals, as this case has a strong
application in practice.


Our problem is motivated by an application arising in VLSI design, where
 one of the main challenges is to build so-called repeater trees. These
are tree-like structures consisting of wires and possibly so-called
repeater circuits and their task is to distribute a signal from a
source circuit to several sink circuits. 
Thereby, the signal is delayed.
In order to guarantee, that the chip works on the desired speed, timing
constraints are given, that is, the signal has to arrive at each sink
circuit not later than a given individual time bound.
There are several heuristics to build such repeater trees (see e.g.
\cite{bartoschek2010repeater}).

A repeater tree can be modeled as a Steiner tree connecting the
source and the sinks and containing repeater circuits at some of the Steiner
points. The length of a repeater tree corresponds to its power consumption. 
So the question arises, if the length of a given tree can be reduced by
moving the positions of the Steiner points.
Bartoschek et al. \cite{bartoschek2010repeater} have shown, that by adding
repeater circuits at appropriate positions the delay of a signal on a path from
the source to a sink is approximately proportional to the length of the path.
Thus the timing constraints directly yield length restrictions on root-terminal
paths.
It turns out, that the \emph{Rectilinear Steiner Tree Problem with given
Topology and Length restrictions} is a good model for the task to minimize the
power consumption of given repeater trees without changing their topology.


If we are allowed to change the topology of the tree, the problem becomes
NP-hard, as it contains the \emph{Rectilinear Steiner Tree Problem}
\cite{garey1977rectilinear}.
If, additionally,  $l_t=||p(s)-p(t)||_1$ for all $t\in T$, that is, all
root-terminal paths are shortest paths, we end at the \emph{Rectilinear Steiner
Arborescence Problem}, which is also NP-hard (\cite{ShiSu,rao1992rectilinear}).
In the case where the length restrictions are the same for all terminals we have
the case of \emph{Shallow Light Steiner Trees}.

However, if we have to keep the topology, but do not have any length
restrictions, an optimal embedding can be computed in linear time using
dynamic programming (see e.g. \cite{JiangWang}).
To our knowledge, the problem of embedding a Steiner tree with a given topology
satisfying length restrictions has not been considered
yet. In this paper we present the first combinatorial polynomial time algorithm
that computes an optimal embedding.

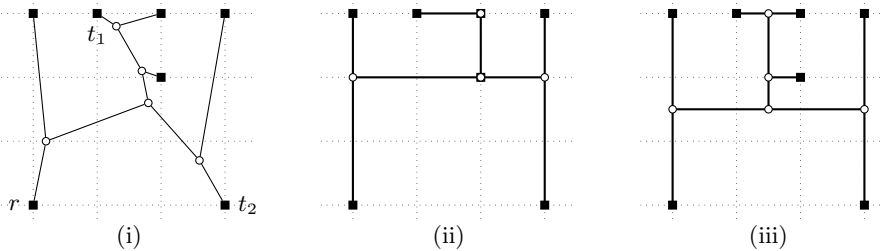
\begin{figure}[ht]
\begin{center}
\begin{tikzpicture}[xscale=0.85,yscale=-0.85]

 \node at ( 1.5,3.5) () {(i)};
 \node at ( 6.5,3.5) () {(ii)};
 \node at (11.5,3.5) () {(iii)};

  \draw[step=1cm, very thin, dotted] (-0.5,-0.25) grid (3.5,3.25);

  \node at (0,3) [terminal, label=left:$r$] (a1) {}; 
  \node at (0,0) [terminal] (a2){}; 
  \node at (2,0) [terminal] (a3) {};
  \node at (2,1) [terminal] (a4) {};
  \node at (3,0) [terminal] (a5) {};
  \node at (1,0) [terminal, label=below:$t_1$] (a6) {};
  \node at (3,3) [terminal, label=right:$t_2$] (a7) {};

  \node at (0.2,2) [circle, draw, fill=white, inner sep=1pt] (b1) {};
  \node at (1.3,0.2) [circle, draw, fill=white, inner sep=1pt] (b2) {};
  \node at (1.7,0.9) [circle, draw, fill=white, inner sep=1pt] (b3) {};
  \node at (1.8,1.4) [circle, draw, fill=white, inner sep=1pt] (b4) {};
  \node at (2.6,2.3) [circle, draw, fill=white, inner sep=1pt] (b5) {};

\draw (a1) -- (b1) -- (a2);
\draw (b1) -- (b4) -- (b3) -- (b2) -- (a3);
\draw (b2) -- (a6);
\draw (b3) -- (a4);
\draw (b4) -- (b5) -- (a7);
\draw (b5) -- (a5);

\begin{scope}[shift={(5,0)}]
 
  \draw[step=1cm, very thin, dotted] (-0.5,-0.25) grid (3.5,3.25);

  \node at (0,3) [terminal] (a1) {}; 
  \node at (0,0) [terminal] (a1){}; 
  \node at (2,0) [terminal] (a1) {};
  \node at (2,1) [terminal] (a1) {};
  \node at (3,0) [terminal] (a1) {};
  \node at (1,0) [terminal] (a1) {};
  \node at (3,3) [terminal] (a1) {};

  \draw[thick] (0,0) -- (0,3);
  \draw[thick] (3,0) -- (3,3);
  \draw[thick] (0,1) -- ++(3,0);
  \draw[thick] (1,0) -- (2,0);
  \draw[thick] (2,0) -- (2,1);

  \node at (0,1) [circle, draw, fill=white, inner sep=1pt] (b1) {};
  \node at (2,0) [circle, draw, fill=white, inner sep=1pt] (b1) {};
  \node at (2,1) [circle, draw, fill=white, inner sep=1pt] (b1) {};
  \node at (2,1) [circle, draw, fill=white, inner sep=1pt] (b1) {};
  \node at (3,1) [circle, draw, fill=white, inner sep=1pt] (b1) {}; 
\end{scope}

\begin{scope}[shift={(10,0)}]
 
  \draw[step=1cm, very thin, dotted] (-0.5,-0.25) grid (3.5,3.25);

  \node at (0,3) [terminal] (a1) {}; 
  \node at (0,0) [terminal] (a1){}; 
  \node at (2,0) [terminal] (a1) {};
  \node at (2,1) [terminal] (a1) {};
  \node at (3,0) [terminal] (a1) {};
  \node at (1,0) [terminal] (a1) {};
  \node at (3,3) [terminal] (a1) {};

  \draw[thick] (0,0) -- (0,3);
  \draw[thick] (3,0) -- (3,3);
  \draw[thick] (0,1.5) -- ++(3,0);
  \draw[thick] (1,0) -- (2,0);
  \draw[thick] (1.5,0) -- (1.5,1.5);
  \draw[thick] (1.5,1) -- (2,1);

  \node at (0,1.5) [circle, draw, fill=white, inner sep=1pt] (b1) {};
  \node at (1.5,0) [circle, draw, fill=white, inner sep=1pt] (b1) {};
  \node at (1.5,1) [circle, draw, fill=white, inner sep=1pt] (b1) {};
  \node at (1.5,1.5) [circle, draw, fill=white, inner sep=1pt] (b1) {};
  \node at (3,1.5) [circle, draw, fill=white, inner sep=1pt] (b1) {}; 
\end{scope}

\end{tikzpicture}
\caption{Instance (i), optimal embedding without length restrictions (ii) 
and optimal embedding with length restrictions $l_{t_1}=5$ and $l_{t_2}=6$ 
(iii). The regular dotted grid has a lattice spacing of $1$.}
\label{fig::example1}
\end{center}
\end{figure}

Figure \ref{fig::example1} (i) shows an instance with
seven terminals drawn as black squares and 5 Steiner points drawn as white
circles.
Figure (ii) shows an optimal solution if there are no length restrictions.
In Figure (iii) an optimal solution is shown, if we have length
restrictions $l_{t_1} = 5$, $l_{t_2}=6$ and $l_{s}=\infty$ otherwise.
If there are no length restrictions, then there always exists an
optimal solution where the Steiner points are positioned at the so called Hanan
grid on $T$ (see \cite{hanan1966steiner}). With length restrictions, this is
no longer true. Nevertheless, we prove  that if the positions of the
terminals and the length restrictions are integral, then there
always exists an solution on half-integral positions.

The problem can be formulated as a linear program by extending the LPs
presented in \cite{cabot1970network,JiangWang}. 
Therefore it can be solved in polynomial time by non-combina\-tori\-al
algorithms.
Nevertheless, we are interested in a combinatorial algorithm for the problem.

After introducing several definitions concerning the movement of components of
the tree in Section \ref{section:component}, we present our main observations in
Section \ref{section:main}. 
Among others, we prove that there always exists an optimal embedding
where the Steiner points are on half-integral positions.
Based on this observation, we introduce in Section \ref{section:dynamic}
a dynamic programming algorithm which is the main ingredient to achieve a
pseudo-polynomial time algorithm.
Refining this algorithm we finally gain a polynomial time algorithm in
Section \ref{section:polynomial}.

\section{Moving Components}
\label{section:component}

Before we come to the main observations of the paper we examine how the 
movements of Steiner points of a given embedding influence the total length of
the tree and the 
length of root-terminal paths. 
First we start with several definitions that we need throughout this paper.

 If $\pi$ is an embedding, then an \emph{x-component} $C$ at \emph{position
$x(C)$} with respect to $\pi$, $x(C)\in\mathbb{R}$, is a
 connected subtree $C$ of $T$  such that all  vertices in
 $C$ have x-coordinate $x(C)$.
 An  x-component $C$ is called \emph{maximal} if there does not exist any
 x-component $C'$ with $C\subsetneq C'$.
A component always depends on the embedding $\pi$. In the following, we omit $\pi$
in the notation if it is clear from the context.
In an analogous way we define a y-component $C$ at position $y(C)$.
In the remainder of the paper we introduce several definitions and state lemmata 
concerning $x$-components. By symmetry, these definitions and lemmata also hold
for $y$-components. 

Let $\Gamma(V(C))$ be the neighbors of the vertices of $C$.
For an x-component $C$ we define 
\begin{eqnarray}
 \Gamma_<^\pi(C) &:=& \{v\in \Gamma(V(C)):\, \pi_x(v) < x(C)\} \text{ and}\\
 \Gamma_>^\pi(C) &:=& \{v\in \Gamma(V(C)):\, \pi_x(v) > x(C)\}.
\end{eqnarray}
In an analogous way we define $\Gamma_<^\pi(C)$ and $\Gamma_>^\pi(C)$  for a
$y$-components $C$. 
If $C$ is a component not containing $r$, then the \emph{predecessor} of $C$ is
the unique vertex $v\in\Gamma_>^\pi(C)\cup\Gamma_<^\pi(C)$ such that $v$ is on
the root-$w$ path for all $w\in V(C)$.
For simplicity of notation we define
\begin{equation}
 \sign(C)=\begin{cases}
           1 & \text{if the predecessor of }C\text{ is in }\Gamma_<(C)\\
           -1& \text{otherwise}.
          \end{cases}
\end{equation}

 If $C$ is an x-component with respect to some embedding $\pi$ then we say that we
 \emph{move $C$  by $\delta$} if we replace $\pi$ by the
 embedding $\pi'$ defined by 
 \begin{equation}
   \pi'(v) :=  
   \begin{cases} \pi(v) + (0,\delta) & \text{for all }v\in V(C)\setminus T, \\
     \pi(v)& \text{otherwise}.
   \end{cases}
 \end{equation}
 We say, that we move $C$ \emph{towards its predecessor} if 
$\delta\cdot\sign(C)<0$.

 If $C$ is a maximal component containing no terminals, then
 we define $R(C)$ to be the set of terminals $t$ such that the unique root-$t$
 path $P$ passes $C$, that is, $V(P)\cap V(C)\neq\emptyset$ and the path enters
and leaves $C$ at the same side, that is,
we have either $|V(P)\cap \Gamma_>(C)|=2$ or $|V(P)\cap \Gamma_<(C)|=2$.
If we choose $\delta\in\mathbb{R}$ with $|\delta|$ small enough and move $C$ by
$\delta$, then the length of all root-$t$ paths with $t\in R(C)$ change by
$2\sign(C)\delta$. The length of any other root-terminal path does not change.

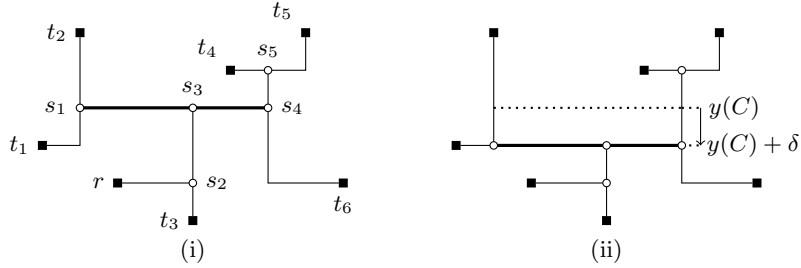
\begin{figure}[ht]
\begin{center}
 \begin{tikzpicture}[scale=0.5]

  \node at (2,1) [terminal, label=left:$r$] (r) {};
  \node at (0,2) [terminal, label=left:$t_1$] (t1) {};
  \node at (1,5) [terminal, label=left:$t_2$] (t2) {};
  \node at (4,0) [terminal, label=left:$t_3$] (t3) {};
  \node at (5,4) [terminal, label=above left:$t_4$] 
(t4) {};
  \node at (7,5) [terminal, label=above left:$t_5$] 
(t5) {};
  \node at (8,1) [terminal, label=below:$t_6$] (t6) {};
    
  \draw (t1) -- ++(1,0) -- (t2);
  \draw[very thick] (1,3) -- (6,3);
  \draw (t6) -- ++(-2,0) -- ++(0,2);
  \draw (r) -- ++(2,0);
  \draw (t3) -- ++(0,3);
  \draw (t4) -- ++(1,0) -- ++(0,-1);
  \draw (t5) -- ++(0,-1) -- ++(-1,0);
    
  \node at (1,3) [circle, draw, fill=white, inner sep=1pt, label=left:$s_1$] 
(s1) {};
  \node at (4,1) [circle, draw, fill=white, inner sep=1pt, label=right:$s_2$] 
(s2) {};
  \node at (4,3) [circle, draw, fill=white, inner sep=1pt, label=above:$s_3$] 
(s3) {};
  \node at (6,3) [circle, draw, fill=white, inner sep=1pt, label=right:$s_4$] 
(s4) {};
  \node at (6,4) [circle, draw, fill=white, inner sep=1pt, label=above:$s_5$] 
(s5) {};

 \node at (4,-0.8) () {(i)};
 
\begin{scope}[xshift=11cm]
  \node at (2,1) [terminal] (r) {};
  \node at (0,2) [terminal] (t1) {};
  \node at (1,5) [terminal] (t2) {};
  \node at (4,0) [terminal] (t3) {};
  \node at (5,4) [terminal] 
(t4) {};
  \node at (7,5) [terminal] 
(t5) {};
  \node at (8,1) [terminal] (t6) {};
    
  \draw (t1) -- ++(1,0) -- (t2);
  \draw[very thick] (1,2) -- (6,2);
  \draw (t6) -- ++(-2,0) -- ++(0,2);
  \draw (r) -- ++(2,0);
  \draw (t3) -- ++(0,2);
  \draw (t4) -- ++(1,0) -- ++(0,-1);
  \draw (t5) -- ++(0,-1) -- ++(-1,0);

  \draw[dotted,thick] (1,3) -- (6,3);
  \draw[dotted,thick] (6,3) -- ++(0.5,0);
  \draw[dotted,thick] (6,2) -- ++(0.5,0);
    
\draw [->] (6.5,3) -- ++(0,-1);
\node at (7.4,3) () {\footnotesize $y(C)$};
\node at (7.9,2) () {\footnotesize $y(C)+\delta$};

  \node at (1,2) [circle, draw, fill=white, inner sep=1pt] 
(s1) {};
  \node at (4,1) [circle, draw, fill=white, inner sep=1pt] 
(s2) {};
  \node at (4,2) [circle, draw, fill=white, inner sep=1pt] 
(s3) {};
  \node at (6,2) [circle, draw, fill=white, inner sep=1pt] 
(s4) {};
  \node at (6,4) [circle, draw, fill=white, inner sep=1pt]
(s5) {};
 
  \node at (4,-0.8) () {(ii)};
 
\end{scope}

 \end{tikzpicture}
 \caption{(i) An embedding $\pi$ with a maximal y-component $C$ with
$V(C)=\{s_1,s_3,s_4\}$, predecessor $s_2$, $\Gamma_>(C)=\{t_2,s_5\}$,
$\Gamma_<(C)=\{t_1,s_2,t_6\}$,
$\sign(C)=1$ and $R(C)=\{t_1,t_6\}$.
 (ii) Embedding obtained by moving $C$ by $\delta<0$. The new embedding
preserves the local positions of $\pi$. The length of all root-$t$ with
$t\in R(C)$ changed by $2\sign(C)\delta=2\delta$.}
 \label{fig:seg}
 \end{center}
\end{figure}

Figure \ref{fig:seg} illustrates some of the definitions.

If $\pi$ and $\pi'$ are two embedding, then we say that \emph{$\pi'$ preserves
the local order of $\pi$} if for every edge $(v,w)\in E(S)$ we have
 \begin{align}
  (\pi_x(v) \leq \pi_x(w))\quad &\Rightarrow\quad 
(\pi'_x(v)\leq\pi'_x(w))\text{ and} \label{eq:implies2}\\
  (\pi_y(v) \leq \pi_y(w))\quad &\Rightarrow\quad 
(\pi'_y(v)\leq\pi'_y(w)).\label{eq:implies4}
 \end{align}
Note, that by (\ref{eq:implies2}) if $\pi_x(v)=\pi_x(w)$ then $\pi'_x(v) =
\pi'_x(w)$ and analogously for $y$.
This implies that each component with
respect to $\pi$ is also a component with respect to $\pi'$ (but not
necessarily the other way round!). 
Moreover, if $v$ is a vertex of an x-component that contain terminals, we
have $\pi_x(v)=\pi'_x(v)$.

Now we can analyze how the length of the embedding and of root-terminal paths
change 
if we move maximal components simultaneously and the local order is preserved.
\begin{lemma}\label{lemma:movement}
 Let $\pi$ be an embedding, $\Delta$  be the set of all maximal x- 
 and y-components, and
 $\delta_C\in\mathbb{R}$ for $C\in\Delta$. Denote by $\pi'$ the 
 embedding we obtain by moving each component $C\in\Delta$ by 
 $\delta_C$.
 If $\pi'$ preserves the local order of $\pi$
 then
 \begin{equation}\label{lemma:eq:lengthchange}
  c(\pi')=c(\pi)+\sum_{C\in\Delta} 
\delta_C\left(|\Gamma^\pi_<(C)| - |\Gamma^\pi_>(C)| 
  \right).
 \end{equation}

 Moreover we have for all $t\in T$:
 \begin{equation}\label{eq:length_root_terminal}
  d_{\pi'}(t) = d_{\pi}(t) + \sum_{C\in\Delta: t\in R(C)} 
2\sign(C) \cdot\delta_C.
 \end{equation}
\end{lemma}
\begin{proof}
 Consider an x-component $C\in\Delta_x$. If we move $C$, then only the length of
edges $\{v,w\}\in E(S)$ with  $v\in V(C)$ and $w\notin  V(C)$ are changed. Let
$\{v,w\}$ be such an edge and assume
$w\in \Gamma^\pi_<(C)$, that is $\pi_x(w) < \pi_x(v)$.
 As the local order is preserved, we have $\pi_x'(w)\leq \pi_x'(v)$. 
But then moving $C$ by $\delta$ increases the length of the edge $\{v,w\}$ by
$\delta$. In an analogous way we see that the length of the edge decreases by
$\delta$ if $w\in\Gamma^\pi_<(C)$.
Summing up the changes over all components we obtain 
(\ref{lemma:eq:lengthchange}).

Now consider a terminal $t\in T$. Again, as the local order is preserved
by
$\pi'$, the length of the root-$t$ path is only influenced by components $C$
with $t\in R(C)$. Consider such a component $C$. If we move $C$ by
$|\delta_C|$ towards the predecessor of $C$, the length of the path is reduced
by $2|\delta_C|$. On the other hand, if we move $C$ in the other direction by
$|\delta_C|$, then the length is increased by $2|\delta_C|$. In total, the
length changes by $\sign(C)2\cdot\delta_C$. Summing up over all such components,
we obtain (\ref{eq:length_root_terminal}). \qed
\end{proof}
The following observation is crucial in order to prove that there exist
optimal solutions that are half-integral.

\begin{figure}[ht]
 \begin{center}
  \begin{tikzpicture}[scale = 0.5]

\node at (3,-1) () {$R(C_1)\cap R(C_2)=\emptyset$};
\node at (3+8,-1) () {$R(C_1)\cap R(C_2)=\emptyset$};
\node at (3+16,-1) () {$R(C_2)\subseteq R(C_1)$};

    \node[draw, fill, inner sep = 1, label=left:$r$]  at (0,0) () {};
    \node at (0.5,3) () {$C_1$};
    \node at (2.5,1) () {$C_2$};
    \draw[very thick] (1,3) -- ++ (3,0);
    \draw[very thick] (3,1) -- ++ (3,0);

    \draw (2.5,3) -- ++(0,0.5);
    \draw (3.5,3) -- ++(0,-0.5);

    \draw (4.5,1) -- ++(0,0.5);
    \draw (5.5,1) -- ++(0,-0.5);

    \draw plot [smooth, tension=1] coordinates { (0,0) (1,1) (1.5,3)};
    \draw plot [smooth, tension=1] coordinates { (1,1) (2.5,0) (3.5,1)};

    \node[draw, circle, fill=white, inner sep = 1, label=above:$v_1$]  at
(1.5,3) () {};
    \node[draw, circle, fill=white, inner sep = 1, label=above:$v_2$]  at
(3.5,1) () {};

    \begin{scope}[shift={(8,0)}]
     \node[draw, fill, inner sep = 1, label=left:$r$]  at (0,0) () {};
     \draw[very thick] (1,1) -- ++(3,0);
     \draw[very thick] (3,3) -- ++(3,0);
     \node at (0.5,1) () {$C_1$};
     \node at (2.5,3) () {$C_2$};
     \draw (3.5,1) -- ++(0,-0.5);
     \draw (4.6,3) -- ++(0,0.5);
     \draw (5.3,3) -- ++(0,-0.5);
     \draw plot [smooth, tension=1] coordinates { (0,0) (1,0.3) (1.5,1)};
     \draw plot [smooth, tension=1] coordinates { (2.4,1) (2.6,1.7) (3.4,2.3)
(3.6,3)};

\draw (2.6,1.7) -- ++(0.6,0);

    \node[draw, circle, fill=white, inner sep = 1, label=above:$v_1$]  at
(1.5,1) () {};
    \node[draw, circle, fill=white, inner sep = 1, label=above:$v_2$]  at
(3.6,3) () {};

    \node[draw, circle, fill=white, inner sep = 1, label=above:$v$]  at
(2.6,1.7) () {};

    \end{scope}

    \begin{scope}[shift={(16,0)}]
     \node[draw, fill, inner sep = 1, label=left:$r$]  at (0,0) () {};

     \draw[very thick] (1,3) -- ++(3,0);
     \draw[very thick] (3,1) -- ++(3,0);
     \node at (0.5,3) () {$C_1$};
     \node at (2.5,1) () {$C_2$};

     \draw (2.4,3) -- ++(0,0.5);
     \draw (4.5,1) -- ++(0,-0.5);
     \draw (5.5,1) -- ++(0,0.5);

     \draw plot [smooth, tension=1] coordinates { (0,0) (0.3,1) (1.3,2.5)
(1.5,3)};
     \draw plot [smooth, tension=1] coordinates { (3.4,3) (3.4,2.3)
(3.7,1.7) (3.7,1)};
    
    \draw (3.4,2.3) -- ++(0.6,0);
    
    \node[draw, circle, fill=white, inner sep = 1, label=above:$v_1$]  at
(1.5,3) () {};
    \node[draw, circle, fill=white, inner sep = 1, label=below:$v_2$]  at
(3.7,1) () {};
    \node[draw, circle, fill=white, inner sep = 1, label=left:$v$]  at
(3.4,2.3) () {};

    \end{scope}

  \end{tikzpicture}

 \end{center}
 \caption{The possible positions of two maximal $y$-components within a tree.}
 \label{figure:twomaxcop}
\end{figure}
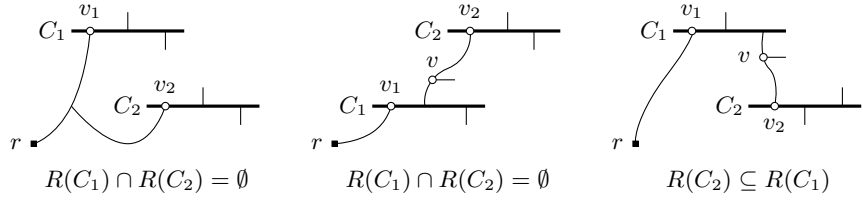

\begin{lemma} \label{lemma:laminar}
 If $\Delta$ is a set of maximal x-components that do not contain terminals,
 then $\{R(C)\}_{C\in \Delta}$ is a laminar family.
\end{lemma}
\begin{proof}
 Let $C_1,C_2\in\Delta$. By definition $V(C_1)\cap V(C_2) = \emptyset$. 
 For $i\in\{1,2\}$ let $v_i$ be the vertex of $V(C_i)$ that is adjacent to the
 predecessor of $C_i$. Note that $v_i$ is on the unique $r$-$t$ path for every
 $t\in R(C_i)$ (see Figure \ref{figure:twomaxcop}).
 Now assume that neither $v_1$ is on the $r$-$v_2$ path nor $v_2$ is on the
 $r$-$v_1$ path.
 Then $R(C_1)\cap R(C_2)=\emptyset$.
 Otherwise, assume w.l.o.g. that $v_1$ is on the $r$-$v_2$ path.
 In this case there exists a unique vertex $v$ on the $v_1$-$v_2$ path satisfying
 $v\in\Gamma^\pi_>(C_1)\cup\Gamma^\pi_<(C_1)$. 
 Now note that the length of all $r$-$t$ paths change for all $t\in R(C_2)$ 
 when moving $C_1$ if and only if the length of the root-$t$ path changes when
 moving $C_1$.
 Hence, $R(C_2)\subseteq R(C_1)$ or $R(C_2)\cap R(C_1)=\emptyset$.
 This implies the desired result. \qed
\end{proof}

Before we continue with the main result we make another simple observation:

\begin{proposition}\label{prop:shortest}
 If $\pi$ is an embedding and there exists a vertex $t\in T$ such that 
 $d_\pi(t) > ||p(v)-p(r)||_1$,
 then there exists a component $C$ such that 
 moving $C$ towards its predecessor decreases the length of the root-$t$ path.
\end{proposition}

\endproof

\section{Main section}
\label{section:main}

In this section we prove that if all terminals are on integral coordinates and
all length restrictions are integral, then
there exists an optimal half-integral embedding. 
More precisely we prove that for any given feasible embedding $\pi$
there exists a feasible half-integral embedding $\sigma$ of at most the same
cost such that the $\ell_\infty$ distance between the positions of a vertex in
both
embeddings  is at most $0.5$.
To this end we consider a sub problem that can be formulated as a linear
program based on a totally unimodular matrix.

We start with some observations on half-integral embeddings.
\begin{proposition}
  Every half-integral embedding has half-integral cost. 
\end{proposition}
\begin{proof}
 Obviously, all edges in such an embedding have half-integral lengths
 and thus the total length is also half-integral. \qed
\end{proof}

\begin{proposition}\label{prop:pathlength}
  In every half-integral embedding $\pi$ the length of every root-terminal path has
integral length.
\end{proposition}
\begin{proof}
 Let $t\in T$ and denote by $P$ the unique root-$t$ path in $S$.
 If $P$ is a shortest path, then the length of $P$ is
 $||\pi(r)-\pi(t)||_1$, which is integral.
 If $P$ is not a shortest path, then by Proposition \ref{prop:shortest} there
 exists a component $C$ such that moving $C$ towards its predecessor decreases
 the length of $P$. As $\pi$ is half-integral, we can move $C$ by $0.5$ towards
 its predecessor, reducing the length of $P$ by $1$ and obtaining a new
 half-integral embedding $\pi'$.
 Then by induction the length of $P$ must be integral. \qed
\end{proof}

The main theorem of this section is the following.
\begin{theorem} \label{lemma:half_move}
 If $\pi$ is an embedding for an integral instance $(S,T,r,p,l)$, then there
 exists an half-integral embedding $\sigma$ with $\max_{v\in V}
 ||\pi(v)-\sigma(v)||_{\infty}
 \leq 0.5$ and $c(\sigma) \leq c(\pi)$.
\end{theorem}
\begin{proof}
 For $x\in \mathbb{R}$ we denote by $I(x)$ the smallest interval in
 $\mathbb{R}$ with half-integral 
 boundaries such that $x$ is in the interior of the interval, that is,
 \begin{equation}
   I(x) := \left[ \lceil 2x -1 \rceil/2, \lfloor 2x+1\rfloor/2 \right].
 \end{equation}
 For a point $(x,y)\in\mathbb{R}^2$ we set
 $
   I((x,y)):= I(x)\times I(y).
 $
 We show that there exists an half-integral embedding $\sigma$ with 
 \begin{equation}\label{eq:corridor}
   \sigma(v)\in I(\pi(v)) \text{ for all } v\in V 
 \end{equation}
 such that $c(\sigma)\leq c(\pi)$.
 
 Let $\sigma$ be a feasible embedding for $(S,T,r,p,l)$ of minimum cost
satisfying 
 (\ref{eq:corridor}). If there are several such embeddings we 
 choose one with a minimal number of components that are not on half-integral
 coordinates.
 We denote this number by $N(\sigma)$ and prove that $N(\sigma)=0$. 
 Suppose that this is not the case.
 The idea is to move maximal components such that
 $N(\sigma)$ gets smaller without increasing $c(\sigma)$.
 As $\pi$ satisfies (\ref{eq:corridor}), we have $c(\sigma) \leq c(\pi)$.

 Let $\Delta_x$ and $\Delta_y$ be the sets of maximal x- and y-components,
 respectively, with respect to $\sigma$  that are not on half-integral
 coordinates and set
 $\Delta:=\Delta_x\dot{\cup} \Delta_y$. Then $N(\sigma) = |\Delta|$.
 For $C\in\Delta$ we set
 \begin{equation}
   z^*_C:= 
   \begin{cases} 
     x(C)- \left\lfloor 2x(C)\right\rfloor/2 & C\in \Delta_x,\\
     y(C)- \left\lfloor 2y(C)\right\rfloor/2 & C\in  \Delta_y.
   \end{cases}
 \end{equation}

 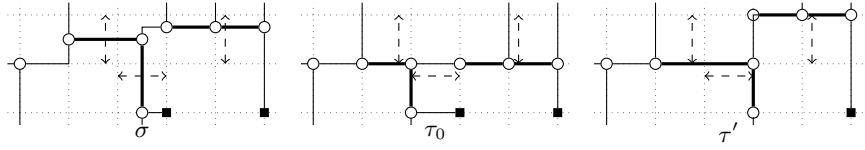
\begin{figure}[ht]
\begin{center}

  \begin{tikzpicture}[scale=0.65,rotate=90, yscale=-1]
    \node at (-1.4,2.5) {$\sigma$};
    \node at (-1.4,8.5) {$\tau_0$};
    \node at (-1.4,14.5) {$\tau'$};

    \draw[step=1cm, very thin, dotted] (-1.25,-0.25) grid (1.25,5.25);

    \node at (0   ,0  ) [draw, circle, fill=white, inner sep=1.5pt] (a1) {};
    \node at (0.5 ,1  ) [draw, circle, fill=white, inner sep=1.5pt] (a2) {};
    \node at (0.5 ,2.5) [draw, circle, fill=white, inner sep=1.5pt] (a3) {};
    \node at (0.75,4  ) [draw, circle, fill=white, inner sep=1.5pt] (a4) {};
    \node at (0.75,5  ) [draw, circle, fill=white, inner sep=1.5pt] (a5) {};
    \node at (0.75,3  ) [draw, circle, fill=white, inner sep=1.5pt] (a7) {};

    \node at (-1,2.5) [draw, circle, fill=white, inner sep=1.5pt] (a6) {};

    \node at (-1,3) [terminal] (t1) {};
    \node at (-1,5) [terminal] (t2) {};

   \draw (a1) -- ++(0,1) -- (a2);
   \draw[very thick] (a2) -- (a3);
   \draw (a3) -- ++(0.25,0) -- (a7);
   \draw[very thick] (a4) -- (a7);
   \draw (a7) -- ++(0.5,0);

   \draw[very thick] (a4) -- (a5);
   \draw (a1) -- ++(-1.25,0);
   \draw (a1) -- ++(0,-0.25);
   \draw (a2) -- ++(0.75,0);
   \draw[very thick] (a3) -- (a6);
   \draw (a6) -- (t1);
   \draw (a6) -- ++(-0.25,0);
   \draw (a4) -- ++(0.5,0);
   \draw (a5) -- ++(0.5,0);
   \draw (t2) -- (a5);

   \draw[dashed, <->] (0,1.75) -- ++(1,0);

   \draw[dashed, <->] (0.75-1,2) -- ++(0,1);
   \draw[dashed, <->] (0,4.2) -- ++(1,0);

\begin{scope}[shift={(0,6)}]
    \draw[step=1cm, very thin, dotted] (-1.25,-0.25) grid (1.25,5.25);

    \node at (0   ,0  ) [draw, circle, fill=white, inner sep=1.5pt] (a1) {};
    \node at (0 ,1  ) [draw, circle, fill=white, inner sep=1.5pt] (a2) {};
    \node at (0 ,2) [draw, circle, fill=white, inner sep=1.5pt] (a3) {};
    \node at (0,4  ) [draw, circle, fill=white, inner sep=1.5pt] (a4) {};
    \node at (0,5  ) [draw, circle, fill=white, inner sep=1.5pt] (a5) {};
    \node at (-1,2) [draw, circle, fill=white, inner sep=1.5pt] (a6) {};
    \node at (-1,3) [terminal] (t1) {};
    \node at (-1,5) [terminal] (t2) {};
    \node at (0,3  ) [draw, circle, fill=white, inner sep=1.5pt] (a7) {};

   \draw (a1)  -- (a2);
   \draw[very thick] (a2) -- (a3);
   \draw[very thick] (a4) -- (a7);
   \draw (a3)  -- (a7) -- (a4);
   \draw[very thick] (a4) -- (a5);
   \draw (a1) -- ++(-1.25,0);
   \draw (a1) -- ++(0,-0.25);
   \draw (a2) -- ++(1.25,0);
   \draw[very thick] (a3) -- (a6);
   \draw (a6) -- (t1);
   \draw (a6) -- ++(-0.25,0);
   \draw (a4) -- ++(1.25,0);
   \draw (a5) -- ++(1.25,0);
   \draw (t2) -- (a5);

   \draw[dashed, <->] (0,1.75) -- ++(1,0);

   \draw[dashed, <->] (0.75-1,2) -- ++(0,1);
   \draw[dashed, <->] (0,4.2) -- ++(1,0);
 
\end{scope}

\begin{scope}[shift={(0,12)}]
    \draw[step=1cm, very thin, dotted] (-1.25,-0.25) grid (1.25,5.25);

    \node at (-1,3) [terminal] (t1) {};
    \node at (-1,5) [terminal] (t2) {};

    \node at (0   ,0  ) [draw, circle, fill=white, inner sep=1.5pt] (a1) {};
    \node at (0 ,1  ) [draw, circle, fill=white, inner sep=1.5pt] (a2) {};
    \node at (0 ,3) [draw, circle, fill=white, inner sep=1.5pt] (a3) {};
    \node at (1,4  ) [draw, circle, fill=white, inner sep=1.5pt] (a4) {};
    \node at (1,5  ) [draw, circle, fill=white, inner sep=1.5pt] (a5) {};

    \node at (-1,3) [draw, circle, fill=white, inner sep=1.5pt] (a6) {};
    \node at (1,3  ) [draw, circle, fill=white, inner sep=1.5pt] (a7) {};

   \draw (a1)  -- (a2);
   \draw[very thick] (a2) -- (a3);
   \draw[very thick] (a4) -- (a7);
   \draw (a3) -- ++(1,0) -- (a7) -- (a4);
   \draw[very thick] (a4) -- (a5);
   \draw (a1) -- ++(-1.25,0);
   \draw (a1) -- ++(0,-0.25);
   \draw (a2) -- ++(1.25,0);
   \draw[very thick] (a3) -- (a6);
   \draw (a6) -- (t1);
   \draw (a6) -- ++(-0.25,0);
   \draw (a4) -- ++(0.25,0);
   \draw (a5) -- ++(0.25,0);
   \draw (t2) -- (a5);

   \draw[dashed, <->] (0,1.75) -- ++(1,0);
   \draw[dashed, <->] (0.75-1,2) -- ++(0,1);
   \draw[dashed, <->] (0,4.2) -- ++(1,0);
 
\end{scope}

  \end{tikzpicture}
\end{center}

  \caption{Detail of an embedding $\sigma$ with three maximal
components not  on half-integral positions. The embedding $\tau'$ preserves the
local order of $\sigma$.
  }
 \label{figure:embedding}
 \end{figure}
 
 Consider a vector $z\in[0,0.5]^\Delta$. Starting with the embedding $\sigma$
and 
 moving each component $C\in\Delta$ by $z_C-z^*_C$ we obtain a new embedding
$\tau(z)$. 
 Note that by the definition of $z^*_C$ this  embedding is half-integral if and 
 only if $z\in \{0,0.5\}^\Delta$.
 Observe that $\tau(0)$ is half-integral, but it does not necessarily satisfy 
 the length restrictions.
 Since by construction $\tau(0)$ preserves the local order of $\sigma$ we
can
 apply Lemma \ref{lemma:movement} and conclude that for all $t\in T$ the
 length of the root-$t$  path with respect to $\tau(0)$ is
 \begin{equation}\label{eq:length_integral}
  d_{\tau(0)}(t) = d_{\sigma}(t) +
\sum_{C\in\Delta:t\in R(C)}2\sign(C)\cdot (-z^*_C).
 \end{equation}
 As $\tau(0)$ is integral, this length is also integral by Proposition
\ref{prop:pathlength}.
 
 Using $z$ as a variable we can formulate a linear program reflecting the new
 cost of the embedding $\tau(z)$ and the length restrictions, under the
assumption that $\tau(z)$
 preserves the local oder of $\sigma$:

 \begin{align}
   & \min  c(\sigma) + \sum_{C\in\Delta} (z_C-z^*_C)\cdot  
      (|\Gamma_<^\pi(C)| - |\Gamma_>^\pi(C)|)   ,    \nonumber \\
    \textrm{s.t. } &  d_{\sigma}(t) + 
     \sum_{C\in \Delta: t\in R(C) }  2\sign(C)(z_C-z^*_C) \leq l_t 
 && \forall t\in T \label{lp1:ineq} \\
    \text{and }& 0\leq 2z_C \leq 1 && \forall C\in\Delta.
 \end{align}
 As $z=z^*$ is a feasible solution the linear program has an optimal solution
 (see also Figure \ref{figure:embedding}).
 Substituting $2z_C$ by $z'_C$ for all $C\in \Delta$ and using (\ref{eq:length_integral})
 we obtain the modified linear program (P'):

    \begin{align}
  &  \min  \sum_{C\in\Delta} z'_C/2\cdot (|\Gamma_<^\pi(C)| - |\Gamma_>^\pi(C)|),   \nonumber \\
    \textrm{s.t.}  &   \sum_{C\in \Delta: t\in R(C) }  \sign(C)z'_C \leq l_t 
        - d_{\tau(0)}(t) && \forall t\in T \label{lp::ineq}\\
    \text{and }& 0\leq z'_C \leq 1 && \forall C\in\Delta. \label{lp::ineq2}
    \end{align}
    
 We show that the matrix $A$ defined by the left side of the inequalities
 (\ref{lp::ineq}) is totally unimodular.
 Note that all entries of a column of $A$  are either non-negative or
non-positive.
 Thus multiplying all rows with non-positive entries by $-1$ we obtain a
 non-negative matrix where each column correspond to the characteristic vectors
 of $\{R(C)\}_{C\in \Delta}= \{R(C)\}_{C\in \Delta_x} \dot{\cup} \{R(C)\}_{C\in
 \Delta_y}$.
 Recall, that by Lemma \ref{lemma:laminar} the sets $\{R(C)\}_{C\in
 \Delta_x}$ and $\{R(C)\}_{C\in \Delta_y}$ are laminar families. We conclude
 that the rows of $A$ correspond to the characteristic vectors of the union of
 two laminar families.
 Edmonds \cite{edmonds1970} proved, that such matrices are totally unimodular.

 Consequently, as the right hand side of  (\ref{lp::ineq}) is integral,
 the constraints in (\ref{lp::ineq2}) are integral and
 $A$ is totally unimodular, there exists an optimal solution
 for (P') that is integral which further implies that
 the original LP has an half-integral optimal solution $\hat{z}$.
 But then $\tau(\hat{z})$ is also half integral and satisfies
 (\ref{eq:corridor}) and  $c(\tau(\hat{z})) \leq c(\sigma)$.

If $\tau(\hat{z})$ preserves the local order of $\sigma$, then $\tau(\hat{z})$
is the embedding we are looking for.
Otherwise choose $\lambda>0$ minimal such that $\tau_\lambda$
defined by $\tau_\lambda(v)=\lambda \sigma(v) + (1-\lambda) \tau(\hat{z})$
preserves the local order of $\sigma$. As the cost and length functions are
convex, $\tau_\lambda$ is a feasible embedding and $c(\tau_\lambda)\leq \lambda
c(\sigma)+(1-\lambda) c(\tau(\hat{z}))\leq c(\sigma)$. Moreover, every maximal
component of $\sigma$ is a component of
$\tau_\lambda$. If $N(\tau_\lambda)=N(\sigma)$, then $\tau_{\lambda-\epsilon}$
also preserves the local order of $\sigma$ for $\epsilon>0$ small enough,
contradicting the choice of $\lambda$.
Thus $N(\tau_\lambda)<N(\sigma)$ contradicting the choice of $\sigma$.
This finishes the proof.\qed
\end{proof}

\begin{conclusion}\label{main_conclusion}
 If all positions and length restrictions are integral, then there exists an
 optimal embedding that is half-integral.
\end{conclusion}

We can improve a non-optimal half-integral embedding by minor movements of
vertices.

\section{Dynamic programming.}
\label{section:dynamic}

A consequence of the previous section is, that any non-optimal half-integral 
embedding can be improved by small half-integral movements of the Steiner 
points. 
\begin{lemma}\label{lemma:step}
 If $\pi$ is a half-integral embedding that is not optimal, then there exists a 
 half-integral embedding $\pi'$ with $\pi(v)-\pi'(v)\in \{-0.5,0,0.5\}^2$
for all $v\in V(S)$ and $c(\pi') \leq c(\pi)-0.5$.
\end{lemma}
\begin{proof}
 Let $\sigma$ be an optimal half-integral embedding.
 For $\lambda\in (0,1)$ we define $\pi_\lambda$ by $\pi_\lambda(v)=\lambda 
 \pi(v)+(1-\lambda)\sigma(v)$ for all $v\in V(S)$. As $\pi$ is not optimal and 
 by the convexity of the length function, $\pi_\lambda$ is a feasible
 embedding and we have $c(\pi_\lambda)\leq \lambda 
 c(\pi)+(1-\lambda)c(\sigma) \leq c(\pi)$. Choose $\lambda$ small enough such 
 that $\max_{v\in V(S)}||\pi(v)-\pi_\lambda(c)||_\infty < 0.5$.
 Now we can apply Theorem \ref{lemma:half_move} yields a half-integral 
 embedding  $\pi'$ satisfying $\max_{v\in V} || \pi(v)-\pi'(v)||_\infty \leq
 \max_{v\in V} || \pi(v)-\pi_\lambda(v)||_\infty + || 
 \pi_\lambda(v)-\pi'(v)||_\infty < 1$
 and $c(\pi')\leq c(\pi_\lambda) < c(\pi)$.
 The claim follows by observing that $\pi'$ and $\pi$ are half-integral. \qed
\end{proof}

This lemma gives a direct idea for an algorithm based on dynamic programming
to improve a non-optimal half-integral embedding. 
In the following, we interpret $S$ as an arborescence rooted at $r$ and
denote by $\Gamma^+(v)$ the children of a vertex $v\in V(S)$. 
For simplicity of notation we set $\pi_\delta(v):=\pi(v)+\delta$ for $\delta\in
\{-0.5,0,0.5\}$.
Moreover, we expand the definition of length restrictions to Steiner points: 
Initially we set $l^\pi_t :=l_t$ for all $t\in T$.
For each vertex $v\in V(S)$ whose children have a length restriction, we set
$$l_v^\pi= \min_{w\in\Gamma^+(v)} l^\pi_w - ||\pi(v)-\pi(w)||_1.
$$

Given an half-integral embedding $\pi$ we want to computes a half-integral 
embedding $\pi'$ with $\pi(v)-\pi'(v)\in\{-0.5,0,0.5\}^2$ and $c(\pi')$ minimal.
Note, that in this case the length of every root-terminal path changes by at
most $2n$.
As, additionally, $\pi'$ is half-integral, 
$l_v^{\pi'}$ is half-integral and $|l_v^{\pi'} - l_v^\pi| \leq 2n$ for all $v\in
V(S)$.

Thus it is sufficient to compute for every vertex $v\in V(S)$,
every translation $\delta\in \{-0.5,0,0.5\}^2$ and every possible
length restriction $l\in \{l^\pi_v-2n,l^\pi_v-2n+0.5,\ldots, l^\pi_v-2n+2n-0.5, l^\pi_v+2n\}$
the minimum length $\gamma(v,\delta,l)$ of an embedding of the arborescence rooted at $v$
such that $v$ is positioned at $\pi_\delta(v)$ and $v$ satisfies the length
restriction $l$.
For a terminal $t$ we have
$\gamma(t,\delta,l)=0$ if $\delta=(0,0)$ and $l\leq l_t$. Otherwise, we set
$\gamma(t,\delta,l)=\infty$.
For all other vertices $v\in V(T)$ we obviously have
$\gamma(v,\delta,l)=$
\begin{equation*}
\sum_{w\in\Gamma^+(v)}
\min_{\delta'\in \{-0.5,0,0.5\}^2}
\gamma\left(w,\delta',l-||\pi_\delta(v)-\pi_{\delta'}(w)||_1\right) +
||\pi_\delta(v)-\pi_{\delta'}(w)||_1.
\end{equation*}

It follows, that the length of an optimal embedding $\pi'$ with $\pi(v)-\pi'(v)\in
 \{-0.5,0.0.5\}^2$
is $\gamma(r,(0,0),0)$.
This number can be computed in $O(n^2)$ time: There are $O(n^2)$ different
triples $(v,\delta,l)$ for which $\gamma(v,\delta,l)$ has to be computed and
each of these computations can be done in constant time.

To compute a global optimal solution, we start with the trivial embedding, where all Steiner
points are positioned at the root. This solution has cost $C=\sum_{t\in T} ||p(t)||_1$.
Then we apply the dynamic programming approach as long as the cost of the
newly computed embedding decrease. 
As the cost is reduced by at least $0.5$ in every round,  we must obtain an
optimal embedding after $2C$ iterations.
Thus our algorithm has a pseudo polynomial running time of $O(Cn^2)$.
In the next section we show how to refine this approach in 
order to achieve a polynomial running time.


\section{An optimal polynomial time algorithm}
\label{section:polynomial}

We refine the ideas of the previous sections in order to obtain a polynomial
time algorithm for our problem.
In the first algorithm the Steiner points are moved by at most $0.5$ in each
direction in every call of the dynamic programming.
The idea of the refined algorithm is to move the Steiner points by $2^k$ for
a suitable $k\in\mathbb{Z}$ in the first rounds. As soon as no improvements
can be obtained by moving Steiner points by $2^k$, we reduce the moving distance
to $2^{k-1}$ and continue applying the dynamic programming.
Repeating this procedure we finally move the Steiner points by $0.5$, obtaining an
optimal embedding. To prove the polynomial running time and to apply the
results of the previous sections we have to consider slightly modified
instances where all terminals are on $2^k$-integral positions.
Here a number $x\in\mathbb{R}$ is called \emph{$2^k$-integral} if
$x/2^k\in\mathbb{Z}$.

First we state a trivial lemma on the existence of feasible embeddings.
\begin{lemma}\label{lemma:feasible}
 There exists a feasible embedding (and thus an optimal one) for $(S,T,r,p,l)$
if
 and only if $||p(t)||_1 \leq l_t$ for all $t\in T$.
\end{lemma}
\begin{proof}
 If there exists a feasible embedding, then obviously $||p(v)||_1 
 \leq l(v)$ for all $v\in T$. If on the other hand $||p(v)||_1 \leq l(v)$, then
 placing all internal vertices on the position of the root is a feasible embedding 
 satisfying the length restrictions. \qed
\end{proof}

For each $k\in\mathbb{N}$ we define a new instance $I_k:=(S,T,r,p_k,l_k)$ on the
same set of terminals and the same topology, but with new positions 
\begin{equation*}
 p_k(v) = (2^k \lfloor p_x(v)/2^k\rfloor, 2^k \lfloor p_y(v)/2^k\rfloor)
\end{equation*}
for all $v\in T$ and length restrictions 
\begin{equation*}
 l_k(v) = 2^k \lfloor (l(v)- ||p(v)-p_k(v) ||_1)/2^k \rfloor
\end{equation*}
for $v\in T$.
In other words we move each terminal towards the root onto the next
$2^k$-integral position and round each length restriction to the next lower
multiple of $2^k$.

If there exists a feasible embedding for $(S,T.r,p,l)$, then there exists also a
feasible embedding for $(S,T,r,p_k,l_k)$: To show this it is sufficient to prove
that $||p_k(v)||_1 \leq l_k(v)$ for all $v\in T$ by Lemma \ref{lemma:feasible}. 
By the choice of $p_k$ and $l_k$ we have
\begin{equation}
  ||p_k(v)||_1 = ||p(v)||_1 - ||p(v)-p_k(v)||_1 \leq l(v) - ||p(v)-p_k(v)||_1 
  \leq l_k(v).
\end{equation}

Set $m:=\min \{m\in\mathbb{N}:\, |p_x(v)| < 2^m \text{ and } |p_y(v)| < 2^m\,
\forall v\in V(S) \}$.
Thus $m$ is the smallest $m\in\mathbb{N}$ such that $p_m(v)=(0,0)$ for all
$v\in V(S)\setminus T$.
\begin{remark}\label{remark:pol}
  The number $m$ is polynomially bounded in the size of the instance. 
\end{remark}

In $(S,T,r,p_m,l_m)$ all terminals are placed at the position of the root. Thus 
placing all internal vertices to that position yields a trivial optimal 
solution of length $0$.

Now we compute by induction an optimal embedding for $(S,T,r,p_{k-1},l_{k-1})$
 given an optimal embedding for $(S,T,r,p_k,l_k)$.

As $m$ is polynomially bounded in the size of the input, each iteration 
can be computed in polynomial time and $(S,T,r,p,l)=(S,T,r,p_0,l_0)$, we obtain
a
optimal solution in polynomial time.

\begin{lemma}\label{lemma:costs}
 Denote by $\sigma_k$ an optimal solution for $I_k$ for all $k\in\mathbb{N}$.
 Then for $k\in\mathbb{N}$ we have $c(\sigma_{k+1}) \leq c(\sigma_{k}) 
 + 6n2^{k}$.
\end{lemma}
\begin{proof}
 Starting with $\sigma_{k}$ we construct a feasible embedding for 
 $I_{k+1}$. By Lemma \ref{lemma:half_move} we can assume w.l.o.g. that all 
 internal vertices of $S$ are on $2^{k-1}$-integral positions in $\sigma_k$.
 We define $\pi$ by setting $\pi(t)=p_{k+1}(t)$ for $t\in T$ and 
 $\pi(v)=\sigma_{k}(v)$ for $v\in V(S)\setminus T$. By this setting we have
 \begin{equation*}
  ||\pi(t)-\sigma_k(t)||_1 = ||p_{k+1}(t)-p_{k}(t)||_1\leq 2^{k+1}
 \end{equation*}
 for all $t\in T$.
 Thus 
 \begin{equation}\label{eq:first_add}
   c(\pi) \leq c(\sigma_k) + n2^{k+1}
 \end{equation}
 and the length of each root-terminal path increased by at most $2^{k+1}$.
 As $l_k(t) \leq l_{k+1}(t) + 2^{k+1}$ for all $t\in T$,
 we conclude that for each $t\in T$ the length restriction $l_{k+1}(t)$ is hurt
in $\pi$ by at most $2^{k+2}$:
 \begin{equation}
   \sum_{e\in E[r,t]} \pi(e) \leq \sum_{e\in E[r,t]} \sigma_k(e) + 2^{k+1} \leq 
l_k(t) + 2^{k+1} \leq l_{k+1}(t)+ 2^{k+2}.
 \end{equation}

 Now we move components towards their predecessors, until all length
restrictions are satisfied.
 To this end, denote by $\Delta$ the set of all maximal components, that do not
contain the root $r$. Moving all components $C\in \Delta$ by $2^{k-1}$ towards
its predecessors we obtain a new feasible embedding $\pi'$ with $c(\pi') \leq
c(\pi)+n2^k$. 
Moreover, the length of every root-terminal path that has not been a shortest
one with respect to $\pi$ is reduced by at least $2^k$.
Repeating this process with the new embedding at most 3 times yields a feasible
embedding $\pi^*$ for $I_{k+1}$. We conclude $c(\pi^*)\leq c(\pi)+ 4n2^{k}$.
Together with (\ref{eq:first_add}) we conclude $c(\pi^*)\leq
c(\sigma_{k})+6n2^{k}$.
We finish the proof by observing, that as $\pi^*$ is feasible for $I_{k+1}$, the
embedding $\sigma_{k+1}$ cannot be longer. \qed
\end{proof}

Combining the observation of the previous Lemma we obtain our main result. 
\begin{theorem}
 The rectilinear Steiner tree embedding problem with length restrictions can be
 solved in polynomial time by a combinatorial algorithm.
\end{theorem}
\begin{proof}
 Let $I=(S,T,r,p,l)$ be an instance of the problem.
 First we calculate $m$ as above. $m$ is polynomially bounded in the size of the
 input.
 Now we have a polynomial number of instances $I_k$, $k\in\{1,\ldots, m\}$.
 For $I_m$ we have the trivial embedding $\sigma_m$ with $\pi_m(v)=0$ for all
 $v\in V$.
 
 Let $\sigma_{k+1}$ be an optimal embedding for $I_{k+1}$. Then $\pi$ defined as
 $\pi(v)=p_{k}(v)$ for $v\in T$ and $\pi(v)=\sigma_{k+1}(v)$ otherwise is a
 feasible embedding for $I_{k}$. By Lemma \ref{lemma:costs}, 
 $c(\pi) \leq c(\sigma_{k+1}) + n2^{k} \leq
 O_k + 7 n2^{k}$ where $O_k$ denotes the optimal length of an embedding for $I_k$.
  Moreover, $I_{k}$ is an $2^{k}$ integral instance. Thus applying
 the dynamic programming from the previous section at most $14n$ times
 we obtain an optimal solution $\sigma_k$ for $I_k$.
 We conclude that computing $\sigma_k$ from $\sigma_{k+1}$ requires at most
 time $O(n^3)$.
 By induction we get an optimal solution $\sigma$ for $I_1=(S,T,r,c,l)$.
 The total running time is $O(mn^3)$ where 
 $m=\lceil \max \{ |p_x(t)|,|p_y(t)|:\, t\in T\} +1 \rceil$. \qed
\end{proof}

%
%
%

%

\newcommand\drawTerminals{
 \node at (0,0) [terminal] () {};
 \node at (0,4) [terminal] () {};
 \node at (1,5) [terminal] () {};
 \node at (2,6) [terminal] () {};
 \node at (6,7) [terminal] () {};
 \node at (3,7) [terminal] () {};
 \node at (3,1) [terminal] () {};
 \node at (4,-1) [terminal] () {};
 \node at (7,-1) [terminal] () {};
 \node at (9,-1) [terminal] () {};
 \node at (9,2) [terminal] () {};
 \node at (7,2) [terminal] () {};
 \node at (6,2) [terminal] () {};
 \node at (10,1) [terminal] () {};
}

\begin{figure}[ht]
\begin{tikzpicture}[scale=0.5]
\node at (5,-1.5+10) () {(i)};
\node at (13+5,-1.5+10) () {(ii)};

\begin{scope}[shift={(0,10)}]
  \draw[step=1cm, very thin, dotted] (-0.25,-1.25) grid (10.5,7.25);
\node[terminal,label=left:$r$] at (0,0) (a0) {};
\node[terminal] at (0,4) (a1) {};
\node[terminal] at (1,5) (a2) {};
\node[terminal] at (2,6) (a3) {};
\node[terminal] at (6,7) (a4) {};
\node[terminal,label=left:$a$] at (3,7) (a5) {};
\node[terminal,label=left:$b$] at (3,1) (a6) {};
\node[terminal] at (4,-1) (a7) {};
\node[terminal] at (7,-1) (a8) {};
\node[terminal] at (9,-1) (a9) {};
\node[terminal] at (9,2) (a10) {};
\node[terminal,label=above:$c$] at (7,2) (a11) {};
\node[terminal] at (6,2) (a12) {};
\node[terminal] at (10,1) (a13) {};
\node[Steinerpoint] at (2.5,4.5) (a14) {};
\node[Steinerpoint] at (4.5,6) (a15) {};
\node[Steinerpoint] at (4.5,1) (a16) {};
\node[Steinerpoint] at (5.5,0.2) (a17) {};
\node[Steinerpoint] at (7,0) (a18) {};
\node[Steinerpoint] at (8.5,0) (a19) {};
\node[Steinerpoint] at (8,1) (a20) {};
\node[Steinerpoint] at (3.8,4.1) (a21) {};
\node[Steinerpoint] at (1,3) (a22) {};
\node[Steinerpoint] at (5,2) (a23) {};
\node[Steinerpoint] at (1.5,4) (a24) {};
\node[Steinerpoint] at (9,0.5) (a25) {};
\draw (a0) -- (a22);
\draw (a1) -- (a22);
\draw (a22) -- (a24);
\draw (a2) -- (a24);
\draw (a24) -- (a14);
\draw (a3) -- (a14);
\draw (a14) -- (a21);
\draw (a21) -- (a15);
\draw (a15) -- (a4);
\draw (a15) -- (a5);
\draw (a21) -- (a23);
\draw (a23) -- (a16);
\draw (a23) -- (a12);
\draw (a6) -- (a16);
\draw (a16) -- (a17);
\draw (a17) -- (a7);
\draw (a17) -- (a18);
\draw (a18) -- (a8);
\draw (a18) -- (a19);
\draw (a19) -- (a9);
\draw (a19) -- (a25);
\draw (a20) -- (a25);
\draw (a25) -- (a13);
\draw (a20) -- (a10);
\draw (a20) -- (a11);
\end{scope}

\begin{scope}[shift={(13,10)}]
  \draw[step=1cm, very thin, dotted] (-0.25,-1.25) grid (10.5,7.25);
\node[terminal] at (0,0) (a0) {};
\node[terminal] at (0,4) (a1) {};
\node[terminal] at (1,5) (a2) {};
\node[terminal] at (2,6) (a3) {};
\node[terminal] at (6,7) (a4) {};
\node[terminal] at (3,7) (a5) {};
\node[terminal] at (3,1) (a6) {};
\node[terminal] at (4,-1) (a7) {};
\node[terminal] at (7,-1) (a8) {};
\node[terminal] at (9,-1) (a9) {};
\node[terminal] at (9,2) (a10) {};
\node[terminal] at (7,2) (a11) {};
\node[terminal] at (6,2) (a12) {};
\node[terminal] at (10,1) (a13) {};
\node[Steinerpoint] at (2,5) (a14) {};
\node[Steinerpoint] at (4,7) (a15) {};
\node[Steinerpoint] at (4,1) (a16) {};
\node[Steinerpoint] at (4,-1) (a17) {};
\node[Steinerpoint] at (7,-1) (a18) {};
\node[Steinerpoint] at (9,-1) (a19) {};
\node[Steinerpoint] at (9,2) (a20) {};
\node[Steinerpoint] at (4,5) (a21) {};
\node[Steinerpoint] at (0,4) (a22) {};
\node[Steinerpoint] at (4,2) (a23) {};
\node[Steinerpoint] at (1,5) (a24) {};
\node[Steinerpoint] at (9,1) (a25) {};
\draw (a0) -- (a22);
\draw (a1) -- (a22);
\draw (a22) -- (a24);
\draw (a2) -- (a24);
\draw (a24) -- (a14);
\draw (a3) -- (a14);
\draw (a14) -- (a21);
\draw (a21) -- (a15);
\draw (a15) -- (a4);
\draw (a15) -- (a5);
\draw (a21) -- (a23);
\draw (a23) -- (a16);
\draw (a23) -- (a12);
\draw (a6) -- (a16);
\draw (a16) -- (a17);
\draw (a17) -- (a7);
\draw (a17) -- (a18);
\draw (a18) -- (a8);
\draw (a18) -- (a19);
\draw (a19) -- (a9);
\draw (a19) -- (a25);
\draw (a20) -- (a25);
\draw (a25) -- (a13);
\draw (a20) -- (a10);
\draw (a20) -- (a11);
\end{scope}

\end{tikzpicture}
\label{examplerun}
\caption{Instance for the Steiner tree embedding problem (i) and an optimal embedding if there are no length restrictions (ii).}
\end{figure}
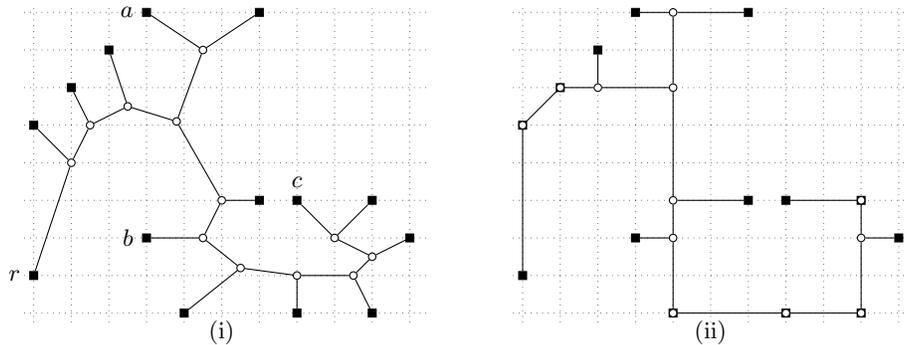

\begin{figure}[ht]
\begin{tikzpicture}[scale=0.33]

\node at (5,-1.5) () {(i)};
\node at (13+5,-1.5) () {(ii)};
\node at (26+5,-1.5) () {(iii)};
\node at (5,-1.5-10) () {(iv)};

\node at (13+5,-1.5-10) () {(v)};
\node at (26+5,-1.5-10) () {(vi)};

\begin{scope}[shift={(0,0)}]
  \draw[step=16cm, very thin, dotted] (-0.25,-1.25) grid (10.5,7.25);
\draw (0,0) -- (0,0);
\draw (0,4) -- (0,0);
\draw (0,0) -- (0,0);
\draw (1,5) -- (0,0);
\draw (0,0) -- (0,0);
\draw (2,6) -- (0,0);
\draw (0,0) -- (0,0);
\draw (0,0) -- (0,0);
\draw (0,0) -- (6,7);
\draw (0,0) -- (3,7);
\draw (0,0) -- (0,0);
\draw (0,0) -- (0,0);
\draw (0,0) -- (6,2);
\draw (3,1) -- (0,0);
\draw (0,0) -- (0,0);
\draw (0,0) -- (4,-1);
\draw (0,0) -- (0,0);
\draw (0,0) -- (7,-1);
\draw (0,0) -- (0,0);
\draw (0,0) -- (9,-1);
\draw (0,0) -- (0,0);
\draw (0,0) -- (0,0);
\draw (0,0) -- (10,1);
\draw (0,0) -- (9,2);
\draw (0,0) -- (7,2);
\drawTerminals
\draw[draw=black, fill=white] (0,0) circle (0.1);
\draw[draw=black, fill=white] (0,0) circle (0.1);
\draw[draw=black, fill=white] (0,0) circle (0.1);
\draw[draw=black, fill=white] (0,0) circle (0.1);
\draw[draw=black, fill=white] (0,0) circle (0.1);
\draw[draw=black, fill=white] (0,0) circle (0.1);
\draw[draw=black, fill=white] (0,0) circle (0.1);
\draw[draw=black, fill=white] (0,0) circle (0.1);
\draw[draw=black, fill=white] (0,0) circle (0.1);
\draw[draw=black, fill=white] (0,0) circle (0.1);
\draw[draw=black, fill=white] (0,0) circle (0.1);
\draw[draw=black, fill=white] (0,0) circle (0.1);

\end{scope}

\begin{scope}[shift={(13,0)}]
  \draw[step=8cm, very thin, dotted] (-0.25,-1.25) grid (10.5,7.25);

\draw (0,0) -- (0,0);
\draw (0,4) -- (0,0);
\draw (0,0) -- (0,0);
\draw (1,5) -- (0,0);
\draw (0,0) -- (0,0);
\draw (2,6) -- (0,0);
\draw (0,0) -- (0,0);
\draw (0,0) -- (0,0);
\draw (0,0) -- (6,7);
\draw (0,0) -- (3,7);
\draw (0,0) -- (0,0);
\draw (0,0) -- (0,0);
\draw (0,0) -- (6,2);
\draw (3,1) -- (0,0);
\draw (0,0) -- (8,0);
\draw (8,0) -- (4,-1);
\draw (8,0) -- (8,0);
\draw (8,0) -- (7,-1);
\draw (8,0) -- (8,0);
\draw (8,0) -- (9,-1);
\draw (8,0) -- (8,0);
\draw (8,0) -- (8,0);
\draw (8,0) -- (10,1);
\draw (8,0) -- (9,2);
\draw (8,0) -- (7,2);
\drawTerminals
\draw[draw=black, fill=white] (0,0) circle (0.1);
\draw[draw=black, fill=white] (0,0) circle (0.1);
\draw[draw=black, fill=white] (0,0) circle (0.1);
\draw[draw=black, fill=white] (8,0) circle (0.1);
\draw[draw=black, fill=white] (8,0) circle (0.1);
\draw[draw=black, fill=white] (8,0) circle (0.1);
\draw[draw=black, fill=white] (8,0) circle (0.1);
\draw[draw=black, fill=white] (0,0) circle (0.1);
\draw[draw=black, fill=white] (0,0) circle (0.1);
\draw[draw=black, fill=white] (0,0) circle (0.1);
\draw[draw=black, fill=white] (0,0) circle (0.1);
\draw[draw=black, fill=white] (8,0) circle (0.1);

\end{scope}
\begin{scope}[shift={(26,0)}]
  \draw[step=4cm, very thin, dotted] (-0.25,-1.25) grid (10.5,7.25);

\draw (0,0) -- (0,4);
\draw (0,4) -- (0,4);
\draw (0,4) -- (0,4);
\draw (1,5) -- (0,4);
\draw (0,4) -- (0,4);
\draw (2,6) -- (0,4);
\draw (0,4) -- (0,4);
\draw (0,4) -- (0,4);
\draw (0,4) -- (6,7);
\draw (0,4) -- (3,7);
\draw (0,4) -- (0,4);
\draw (0,4) -- (0,4);
\draw (0,4) -- (6,2);
\draw (3,1) -- (0,4);
\draw (0,4) -- (4,0);
\draw (4,0) -- (4,-1);
\draw (4,0) -- (8,0);
\draw (8,0) -- (7,-1);
\draw (8,0) -- (8,0);
\draw (8,0) -- (9,-1);
\draw (8,0) -- (8,0);
\draw (8,0) -- (8,0);
\draw (8,0) -- (10,1);
\draw (8,0) -- (9,2);
\draw (8,0) -- (7,2);
\drawTerminals
\draw[draw=black, fill=white] (0,4) circle (0.1);
\draw[draw=black, fill=white] (0,4) circle (0.1);
\draw[draw=black, fill=white] (0,4) circle (0.1);
\draw[draw=black, fill=white] (4,0) circle (0.1);
\draw[draw=black, fill=white] (8,0) circle (0.1);
\draw[draw=black, fill=white] (8,0) circle (0.1);
\draw[draw=black, fill=white] (8,0) circle (0.1);
\draw[draw=black, fill=white] (0,4) circle (0.1);
\draw[draw=black, fill=white] (0,4) circle (0.1);
\draw[draw=black, fill=white] (0,4) circle (0.1);
\draw[draw=black, fill=white] (0,4) circle (0.1);
\draw[draw=black, fill=white] (8,0) circle (0.1);

\end{scope}
\begin{scope}[shift={(0,-10)}]
  \draw[step=2cm, very thin, dotted] (-0.25,-1.25) grid (10.5,7.25);

\draw (0,0) -- (0,4);
\draw (0,4) -- (0,4);
\draw (0,4) -- (2,4);
\draw (1,5) -- (2,4);
\draw (2,4) -- (2,4);
\draw (2,6) -- (2,4);
\draw (2,4) -- (2,4);
\draw (2,4) -- (2,6);
\draw (2,6) -- (6,7);
\draw (2,6) -- (3,7);
\draw (2,4) -- (2,2);
\draw (2,2) -- (2,2);
\draw (2,2) -- (6,2);
\draw (3,1) -- (2,2);
\draw (2,2) -- (4,0);
\draw (4,0) -- (4,-1);
\draw (4,0) -- (8,0);
\draw (8,0) -- (7,-1);
\draw (8,0) -- (8,0);
\draw (8,0) -- (9,-1);
\draw (8,0) -- (8,2);
\draw (8,2) -- (8,2);
\draw (8,2) -- (10,1);
\draw (8,2) -- (9,2);
\draw (8,2) -- (7,2);
\drawTerminals
\draw[draw=black, fill=white] (2,4) circle (0.1);
\draw[draw=black, fill=white] (2,6) circle (0.1);
\draw[draw=black, fill=white] (2,2) circle (0.1);
\draw[draw=black, fill=white] (4,0) circle (0.1);
\draw[draw=black, fill=white] (8,0) circle (0.1);
\draw[draw=black, fill=white] (8,0) circle (0.1);
\draw[draw=black, fill=white] (8,2) circle (0.1);
\draw[draw=black, fill=white] (2,4) circle (0.1);
\draw[draw=black, fill=white] (0,4) circle (0.1);
\draw[draw=black, fill=white] (2,2) circle (0.1);
\draw[draw=black, fill=white] (2,4) circle (0.1);
\draw[draw=black, fill=white] (8,2) circle (0.1);

\end{scope}
\begin{scope}[shift={(13,-10)}]
  \draw[step=1cm, very thin, dotted] (-0.25,-1.25) grid (10.5,7.25);

\draw (0,0) -- (0,4);
\draw (0,4) -- (0,4);
\draw (0,4) -- (1,4);
\draw (1,5) -- (1,4);
\draw (1,4) -- (2,4);
\draw (2,6) -- (2,4);
\draw (2,4) -- (3,4);
\draw (3,4) -- (3,7);
\draw (3,7) -- (6,7);
\draw (3,7) -- (3,7);
\draw (3,4) -- (3,2);
\draw (3,2) -- (3,1);
\draw (3,2) -- (6,2);
\draw (3,1) -- (3,1);
\draw (3,1) -- (4,-1);
\draw (4,-1) -- (4,-1);
\draw (4,-1) -- (7,-1);
\draw (7,-1) -- (7,-1);
\draw (7,-1) -- (7,-1);
\draw (7,-1) -- (9,-1);
\draw (7,-1) -- (7,1);
\draw (7,2) -- (7,1);
\draw (7,1) -- (10,1);
\draw (7,2) -- (9,2);
\draw (7,2) -- (7,2);
\drawTerminals
\draw[draw=black, fill=white] (2,4) circle (0.1);
\draw[draw=black, fill=white] (3,7) circle (0.1);
\draw[draw=black, fill=white] (3,1) circle (0.1);
\draw[draw=black, fill=white] (4,-1) circle (0.1);
\draw[draw=black, fill=white] (7,-1) circle (0.1);
\draw[draw=black, fill=white] (7,-1) circle (0.1);
\draw[draw=black, fill=white] (7,2) circle (0.1);
\draw[draw=black, fill=white] (3,4) circle (0.1);
\draw[draw=black, fill=white] (0,4) circle (0.1);
\draw[draw=black, fill=white] (3,2) circle (0.1);
\draw[draw=black, fill=white] (1,4) circle (0.1);
\draw[draw=black, fill=white] (7,1) circle (0.1);

\end{scope}
\begin{scope}[shift={(26,-10)}]
  \draw[step=0.5cm, very thin, dotted] (-0.25,-1.25) grid (10.5,7.25);

\draw (0,0) -- (0,4);
\draw (0,4) -- (0,4);
\draw (0,4) -- (1,4.5);
\draw (1,5) -- (1,4.5);
\draw (1,4.5) -- (2,4.5);
\draw (2,6) -- (2,4.5);
\draw (2,4.5) -- (3,4.5);
\draw (3,4.5) -- (3,7);
\draw (3,7) -- (6,7);
\draw (3,7) -- (3,7);
\draw (3,4.5) -- (3,2);
\draw (3,2) -- (3,1);
\draw (3,2) -- (6,2);
\draw (3,1) -- (3,1);
\draw (3,1) -- (4,-1);
\draw (4,-1) -- (4,-1);
\draw (4,-1) -- (7,-1);
\draw (7,-1) -- (7,-1);
\draw (7,-1) -- (7,-1);
\draw (7,-1) -- (9,-1);
\draw (7,-1) -- (7,1);
\draw (7,2) -- (7,1);
\draw (7,1) -- (10,1);
\draw (7,2) -- (9,2);
\draw (7,2) -- (7,2);
\drawTerminals
\draw[draw=black, fill=white] (2,4.5) circle (0.1);
\draw[draw=black, fill=white] (3,7) circle (0.1);
\draw[draw=black, fill=white] (3,1) circle (0.1);
\draw[draw=black, fill=white] (4,-1) circle (0.1);
\draw[draw=black, fill=white] (7,-1) circle (0.1);
\draw[draw=black, fill=white] (7,-1) circle (0.1);
\draw[draw=black, fill=white] (7,2) circle (0.1);
\draw[draw=black, fill=white] (3,4.5) circle (0.1);
\draw[draw=black, fill=white] (0,4) circle (0.1);
\draw[draw=black, fill=white] (3,2) circle (0.1);
\draw[draw=black, fill=white] (1,4.5) circle (0.1);
\draw[draw=black, fill=white] (7,1) circle (0.1);
\end{scope}
\end{tikzpicture}
\caption{Run of the algorithm on the instance shown in Figure \ref{examplerun} (i) with 
length restrictions $l_a=10$, $l_b=11$ and $l_c=20$. Figure (vi) shows the final optimal solution.}
\label{examplerun2}
\end{figure}
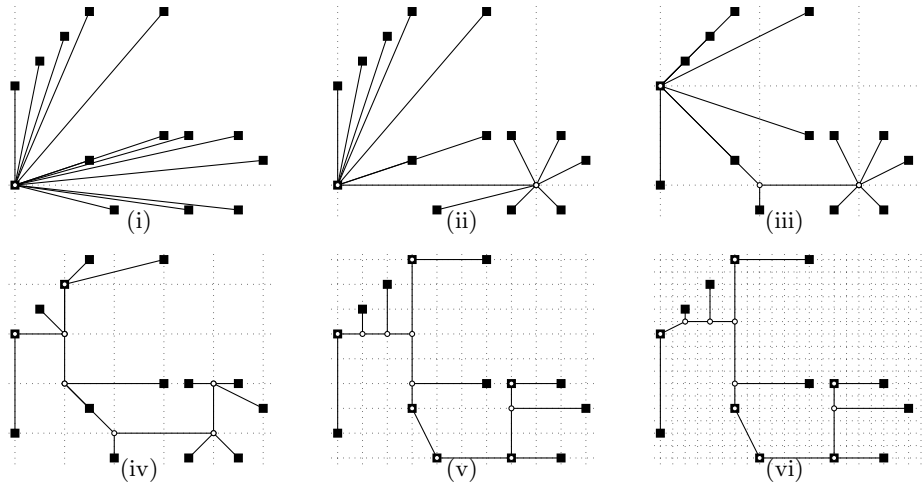
%

Obviously, every feasible solution for $I_k$ corresponds to a feasible solution
for $I$. Moreover, all Steiner points of such embeddings are one
$2^{k-1}$-integral positions.
Due to this observation, the implementation of the algorithm can be modified in
order to decrease the number of dynamic programming steps in practice. Instead
of computing an optimal solution for $I_k$, we are looking for an
embeddings of minimal cost for the original instance $I$, but all Steiner
points have to be on $2^{k-1}$-integral coordinates.
We use the dynamic programming steps as described in order to improve a given
embeddings, but now the cost of each solution is computed
using the original positions of the terminals and consider the original length
restrictions. 
Using this method, the number of dynamic programming steps performed in the
algorithm get very small. It turns out, that in practice, the number of dynamic
programming calls is constant for each $k$ in the most cases.

Figures \ref{examplerun} and \ref{examplerun2} show how the algorithm works on
an example.
Figure \ref{examplerun} (i) shows the instance and Figure \ref{examplerun} (ii)
 an optimal embedding of length $35$ if there are no length restrictions.
In Figure \ref{examplerun2} the embeddings computed by our algorithm are shown.
As input we used the instance from Figure \ref{examplerun} with 
length restrictions  $l_a=10$, $l_b=11$ and $l_c=20$.
As $\max \{|\pi_x(t)|, |\pi_y(t)|:\, t\in T\} = 10$ we have $m=5$.
Thus the algorithm begins with an embedding where all Steiner points are
$2^{m-1}$-integral (Figure (i)). Proceeding with $k=4$ (Figure (ii)) to $k=0$
(Figure (vi)).
The last one is the final optimal embedding of length of $37.5$. For each $k$
the dynamic programming is called at most twice, the first time the length is
reduced, the second time an embedding of the same cost is computed, proving,
that it is an optimal one.

\nocite{*}
\bibliographystyle{plain}
\bibliography{steiner_paper}

\end{document}